\documentstyle[12pt]{article}
\newcommand{\beq}{\begin{equation}}
\newcommand{\eeq}{\end{equation}}
\newcommand{\bea}{\begin{eqnarray*}}
\newcommand{\eea}{\end{eqnarray*}}
\newcommand{\beqa}{\begin{eqnarray}}
\newcommand{\eeqa}{\end{eqnarray}}
\makeatletter
\def\numberbysection{\@addtoreset{equation}{section}
\def\theequation{\thesection.\arabic{equation}}}
\makeatother

\numberbysection
\begin{document}
\centerline{\Large\bf Boundary Action of $N=2$ Super-Liouville Theory}
\vskip 2cm
\centerline{\large Changrim Ahn\footnote{ahn@ewha.ac.kr}
and Masayoshi Yamamoto\footnote{yamamoto@dante.ewha.ac.kr}}
\vskip 1cm
\centerline{\it Department of Physics}
\centerline{\it Ewha Womans University}
\centerline{\it Seoul 120-750, Korea}
\centerline{\small PACS: 11.25.Hf, 11.55.Ds}
\vskip 1cm
\centerline{\bf Abstract}
We derive a boundary action of $N=2$ super-Liouville theory which
preserves both $N=2$ supersymmetry and conformal symmetry
by imposing explicitly $T={\overline T}$ and $G={\overline G}$.
The resulting boundary action shows a new duality symmetry.


\section{Introduction}
Two-dimensional Liouville field theory (LFT) has been studied actively
for its relevance with non-critical string theories and two-dimensional
quantum gravity \cite{LFT,CurTho}.
This theory has been extended to the supersymmetric 
Liouville field theories (SLFTs) which can describe the non-critical 
superstring theories.
In particular, the $N=2$ SLFT 
has been studied actively because the world sheet supersymmetry
can generate the space-time supersymmetry.
Besides applications to the string theories, 
these models provide theoretically challenging problems.
The Liouville theory and its supersymmetric generalizations are 
irrational conformal field theories (CFTs) which have continuously
infinite number of primary fields. 
Due to this property, most CFT formalisms developed for rational CFTs 
do not apply to this class of models.
An interesting problem is to extend the conventional CFT formalism 
to irrational CFTs.
There has been a lot of progress in this field.
Various methods have been proposed to derive structure
constants and reflection amplitudes, which are basic building
blocks to complete the conformal bootstrap \cite{Gervais,Teschner,ZamZam}.
These have been extended to the $N=1$ SLFT in \cite{RasSta,Poghossian}.

More challenging problem is to extend these formalisms to the CFTs
defined in the two-dimensional
space-time geometry with a boundary condition (BC) which preserves
the conformal symmetry.
Cardy showed that the conformally invariant BCs
can be associated with the primary fields in terms of
modular $S$-matrix elements for the case of rational CFTs \cite{Cardy}.
It has been an issue whether the Cardy formalism can be extended to
the irrational CFTs.
There are active efforts to understand the conformally invariant 
boundary states in the context of string theories related to
D-branes \cite{string,DKKMMS}.

An important progress in this direction is made in \cite{FZZ}
where functional relation method
developed in \cite{Teschner} has been used in the boundary LFT.
With a boundary action which preserves the conformal symmetry, 
one-point function of a bulk operator in the presence of the boundary
interaction and two-point correlation functions of boundary operators 
have been computed using the functional relation method \cite{FZZ}.
Here the conformal BC is denoted by a continuous parameter
appearing in the boundary action.
A similar treatment of the LFT defined in the classical Lobachevskiy
plane, namely the pseudosphere has been made in \cite{ZZ}.
For the $N=1$ SLFT, the one-point functions and the boundary 
two-point functions have been obtained in \cite{FH,ARS}
based on the conjectured boundary action.
It is desirable to show that indeed this action preserves
both the supersymmetry and the conformal symmetry.

In this paper we derive the bounary actions of the $N=1,2$ SLFTs
by imposing the symmetries.
This approach to obtain the boundary actions have been made before.
In \cite{W}, based on superfield formulation, the $N=2$ supersymmetric
boundary action has been derived for a general $N=2$ supersymmetric
quantum field theory.
For integrable quantum field theories with infinite conserved charges,
the situation becomes much more complicated.
As shown in a pioneering work \cite{GZ}, the boundary action which
preserves the integrability can be fixed by imposing a first few
conservation laws.
For the supersymmetric integrable models, these two conditions, the
supersymmetry and the integrability, have been successfully imposed
to get appropriate boundary actions \cite{Inami,NN1,NN2}.
We continue this approach to the $N=1,2$ SLFTs and impose
the boundary superconformal invariance conditions to derive
the boundary actions.
We will show that even at classical level, the boundary actions
are determined uniquely.

This paper is organized as follows. In sect.2 we review a superfield
formulation of the $N=1$ SLFT boundary action proposed previously.
Then, we show that this action satisfies the superconformal invariance.
Our main result, the superconformally invariant boundary action of the
$N=2$ SLFT is derived in sect.3.
After repeating the superfield formulation, we derive the boundary 
action by imposing the $N=2$ superconformal symmetry.
We conclude in sect.4 with a few discussions and provide technical 
details in the Appendices.

\section{Boundary $N=1$ Super-Liouville Theory}
In this section, we review a superfield formulation of the
boundary action of the $N=1$ SLFT which 
preserves the boundary $N=1$ supersymmetry.
Then, we will show that the same result can be obtained by
imposing directly the $N=1$ superconformal symmetry.

\subsection{Superfield formulation of the $N=1$ boundary action}

The action of the $N=1$ SLFT is given by \cite{IO}
\beq
S=\int d^2zd^2\theta\left(\frac{1}{2\pi}\bar{D}\Phi D\Phi
+i\mu e^{b\Phi}\right),
\label{N1SLS}
\eeq
where $\Phi$ is a real scalar superfield
\beq
\Phi=\phi+i\theta\psi-i\bar{\theta}\bar{\psi}+i\theta\bar{\theta}F.
\label{Phi}
\eeq
(See the Appendix A.1 for our conventions of the $N=1$ supersymmetry.)
This theory contains a dimensionless Liouville coupling constant $b$
and the cosmological constant $\mu$.
Note that we consider a trivial background and omit a linear 
dilaton coupling.
We can express the action in terms of the component fields
\beq
S=\int d^2z\left[\frac{1}{2\pi}(\partial\phi\bar{\partial}\phi
+\psi\bar{\partial}\psi+\bar{\psi}\partial\bar{\psi})
+i\mu b^2\psi\bar{\psi}e^{b\phi}+\frac{1}{2}\pi\mu^2b^2 e^{2b\phi}\right],
\label{N1SLScomp}
\eeq
by integrating over the $\theta$- and $\bar{\theta}$-coordinates
in (\ref{N1SLS}) and eliminating the auxiliary field $F$ from
its equation of motion.

To introduce the boundary action, we consider
first a general $N=1$ supersymmetric theory on the lower half-plane : 
$-\infty<x={\rm Re}z<\infty,~-\infty<y={\rm Im}z\leq 0$.
Following \cite{DKKMMS}, we can write the action as follows:
\beq
S=\int_{-\infty}^{\infty}dx\int_{-\infty}^0 dy\int d^2\theta{\cal L},
\label{N1S}
\eeq
where ${\cal L}$ is the Lagrangian density in superspace.
The supersymmetry variation of the action is
\beq
\delta S=\int_{-\infty}^{\infty}dx\int_{-\infty}^0 dy\int d^2\theta
(\zeta Q+\bar{\zeta}\bar{Q}){\cal L}
=-\frac{i}{2}\int_{-\infty}^{\infty}dx(\zeta{\cal L}|_{\bar{\theta}}
+\bar{\zeta}{\cal L}|_{\theta})|_{y=0}.
\label{N1dS}
\eeq
To cancel the surface term (\ref{N1dS}), we add a boundary action
\beq
S_B=\frac{i}{2}\eta\int_{-\infty}^{\infty}dx
{\cal L}|_{\theta=\bar{\theta}=0},
~~~\eta=\pm 1,
\label{N1SB}
\eeq
which is defined at $y=0$.
When $\zeta=\eta\bar{\zeta}$,
the supersymmetry variation of the total action vanishes:
$\delta S+\delta S_B=0$.
Only one supercharge 
$Q+\eta\bar{Q}$ is preserved.
Conservation of this charge imposes the boundary condition on 
the supercurrent: $G+\eta\bar{G}=0$ at $y=0$.
The superderivatives in the tangential and normal directions are given by
$D_t=D+\eta\bar{D}$ and $D_n=D-\eta\bar{D}$ respectively.
Their conjugate coordinates are $\theta_t=(\theta+\eta\bar{\theta})/2$
and $\theta_n=(\theta-\eta\bar{\theta})/2$.

For the total variation of $S+S_B$ to vanish, 
two types of boundary conditions can be imposed:

\noindent
(i) Dirichlet boundary conditions $D_t\Phi|_{y=\theta_n=0}=0$:
For the $N=1$ SLFT, this correponds to
\beq
\psi-\eta\bar{\psi}|_{y=0}=0,
~~~\partial_x\phi|_{y=0}=0.
\label{N1SLD}
\eeq
These conditions can be identified with 
the supersymmetric version of the ZZ brane
\cite{ZZ,FH,ARS}

\noindent
(ii) Neumann boundary conditions $D_n\Phi|_{y=\theta_n=0}=0$:
For the $N=1$ SLFT, these give 
\beq
\psi+\eta\bar{\psi}|_{y=0}=0,
~~~\partial_y\phi-2\eta\pi\mu be^{b\phi}|_{y=0}=0.
\label{N1SLN}
\eeq
These boundary conditions correspond to the supersymmetric version of 
the FZZT brane \cite{FZZ,T}.

In \cite{DKKMMS}, it is shown that one can add additional term to
the boundary action
\beq
S'_B=-\frac{1}{2}\int_{-\infty}^{\infty}dx\int d\theta_t
\left(\Gamma D_t\Gamma +\frac{4}{b}i\mu_B\Gamma e^{b\Phi/2}\right).
\label{N1SLSBsuper}
\eeq
with a fermionic boundary superfield $\Gamma=a+i\theta_t h$.
In fact, this boundary action is equivalent to that considered previously
in \cite{ARS,FH}.
We will show in the next subsection that this action indeed preserves
the boundary superconformal symmetry.

\subsection{Boundary superconformal symmetry}

To derive a boundary action which preserves 
both $N=1$ supersymmetry and conformal symmetry, 
we start with a general form of boundary action
\beq
S_B=\int_{-\infty}^{\infty}dx\left[-\frac{i}{4\pi}\bar{\psi}\psi
+\frac{1}{2}a\partial_x a -f(\phi)a(\psi+\bar{\psi})+B(\phi)\right],
\label{N1SLSB}
\eeq
where $a$ is a real fermionic boundary degree of freedom
which anti-commutes with $\psi$ and $\bar{\psi}$.
The boundary action (\ref{N1SLSB}) was first proposed
in the boundary $N=1$ supersymmetric sine-Gordon model \cite{NN1}.
$f(\phi)$ and $B(\phi)$ are functions of the scalar field $\phi$
to be determined by the boundary conditions
which preserve $N=1$ supersymmetry.
The fermionic boundary degree of freedom $a$ was first introduced
in the Ising model in a boundary magnetic field \cite{GZ}
and in the $N=1$ SLFT with appropriate kinetic term \cite{ARS}.

The boundary $N=1$ superconformal symmetry
imposes the following constraints on the stress tensor 
and supercurrent
\beq
T=\bar{T},
~~~G=\bar{G}
~~~{\rm at}
~~~y=0.
\label{N1bc}
\eeq
Here we choose $\eta=-1$ and preserve only one supercharge $Q-{\overline Q}$.
Hence, it is called sometimes as $N=1/2$ supersymmetry.

The stress tensor $T$ and the supercurrent $G$ are given by
\beq
T=-\frac{1}{2}\left((\partial\phi)^2+\psi\partial\psi\right)
+\frac{1}{2}{\widehat Q}\partial^2\phi,\qquad
G=i(\psi\partial\phi-{\widehat Q}\partial\psi),
\label{N1G}
\eeq
where ${\widehat Q}$ is the background charge.
By using the equations of motion,
one can easily show that the conservation laws
$\bar{\partial}T=\partial\bar{T}=\bar{\partial}G=\partial\bar{G}=0$
are satisfied at the classical level with ${\widehat Q}=1/b$.

Using the bulk equations of motion 
\beqa
&&\partial\bar{\partial}\phi=\pi\mu b^3(i\psi\bar{\psi}
+\pi\mu e^{b\phi})e^{b\phi},
\nonumber\\
&&\bar{\partial}\psi=-\pi i\mu b^2\bar{\psi}e^{b\phi},
\qquad\partial\bar{\psi}=\pi i\mu b^2\psi e^{b\phi}.
\label{N1eom}
\eeqa
and the boundary equations of motion 
\beq
\partial_y\phi=4\pi\frac{\partial f}{\partial\phi}a(\psi+\bar{\psi})
-4\pi\frac{\partial B}{\partial\phi},\quad
\psi-\bar{\psi}=-4\pi ifa,\quad\partial_x a=f(\psi+\bar{\psi}),
\label{N1boeom}
\eeq
we obtain
\beq
G-\bar{G}=2\pi\left(f-\frac{2}{b}\frac{\partial f}{\partial\phi}\right)
\partial_x\phi a
+\pi\left(-\frac{2}{f}\frac{\partial B}{\partial\phi}-\frac{4}{b}f
+\mu be^{b\phi}\frac{1}{f}\right)\partial_x a.
\label{N1GbarG}
\eeq
Here we eliminated $\psi,\bar{\psi}$ assuming $f$ is not zero.

The condition $G-\bar{G}=0$ can be satisfied by the following $f$ and $B$
\beq
f=\mu_B e^{b\phi/2},\qquad 
B=\left(-\frac{2}{b^2}\mu_B^2+\frac{1}{2}\mu\right)e^{b\phi},
\label{N1B}
\eeq
where $\mu_B$ is the boundary cosmological constant.
One can show similarly that $T-\bar{T}=0$ can be also satisfied. 
One can easily check that the boundary action (\ref{N1SLSBsuper})
with (\ref{N1SB}) in terms of the superfields
is indeed the same as (\ref{N1SLSB}) with (\ref{N1B}).
Therefore, this action preserves not only boundary $N=1$ supersymmetry
but also conformal symmetry.

So far, we have considered only the classical equations of motion.
Even at this level, the boundary action has been determined uniquely.
We can consider quantum corrections in similar approach.
For this, we interpret $e^{b\phi}$ in Eqs.(\ref{N1eom}) 
as the normal-ordered exponential $:e^{b\phi}:$.
The fields in the stress tensor and the supercurrent in (\ref{N1G})
should be also normal-ordered.
With this change, we obtain
\beq
\bar{\partial}T=\pi\mu b^2(1+b^2-{\widehat Q}b)[\pi\mu\partial
(:e^{b\phi}:)^2+i\psi\bar{\psi}\partial:e^{b\phi}:
-i\bar{\psi}\partial\psi:e^{b\phi}:].
\label{conlaw}
\eeq
The conservation law $\bar{\partial}T=0$ (and others) can be
satisfied when the background charge is renormalized to ${\widehat Q}=1/b+b$.
We will show in Appendix B that the boundary superconformal symmetry
$T-\bar{T}=0$ and $G-\bar{G}=0$ is also preserved at the quantum level
with this ${\widehat Q}$.

\section{Boundary $N=2$ Super-Liouville Theory}
In this section, we use previous method to derive the 
superconformal boundary action of the $N=2$ SLFT.

\subsection{Superfield Formulation}

The action of the $N=2$ SLFT is given by
\beq
S=\int d^2 z\left[\frac{1}{\pi}\int d^4\theta\Phi^+\Phi^-
+\left(i\mu\int d^2\theta^+e^{b\Phi^+}+{\rm c.c.}\right)\right],
\label{N2SLS}
\eeq
where $\Phi^{\pm}$ are the chiral superfields which satisfy
\beq
D_{\mp}\Phi^{\pm}=\bar{D}_{\mp}\Phi^{\pm}=0.
\label{chiral}
\eeq
Therefore, $\Phi^{\pm}$ can be expanded as
\beq
\Phi^{\pm}=\phi^{\pm}(y^{\pm},\bar{y}^{\pm})
+i\theta^{\pm}\psi^{\mp}(y^{\pm},\bar{y}^{\pm})
-i\bar{\theta}^{\pm}\bar{\psi}^{\mp}(y^{\pm},\bar{y}^{\pm})
+i\theta^{\pm}\bar{\theta}^{\pm}F^{\pm}(y^{\pm},\bar{y}^{\pm}),
\label{chiralex}
\eeq
where $y^{\pm}=z+\frac{1}{2}\theta^{\pm}\theta^{\mp}$
and $\bar{y}^{\pm}=\bar{z}+\frac{1}{2}\bar{\theta}^{\pm}\bar{\theta}^{\mp}$.
(See Appendix A.2 for conventions.)
The action can be written in terms of the component fields as
\beqa
S&=&\int d^2z\Bigg[\frac{1}{2\pi}\left(\partial\phi^-\bar{\partial}\phi^+
+\partial\phi^+\bar{\partial}\phi^-
+\psi^-\bar{\partial}\psi^++\psi^+\bar{\partial}\psi^-
+\bar{\psi}^-\partial\bar{\psi}^++\bar{\psi}^+\partial\bar{\psi}^-\right)
\nonumber\\
&&+i\mu b^2\psi^-\bar{\psi}^-e^{b\phi^+}
+i\mu b^2\psi^+\bar{\psi}^+e^{b\phi^-}
+\pi\mu^2 b^2e^{b(\phi^++\phi^-)}\Bigg].
\label{N2L}
\eeqa

Now we consider boundary conditions in the $N=2$ SLFT on the lower half-plane.
The action can be written as
\beqa
S&=&\int_{-\infty}^{\infty}dx\int_{-\infty}^0 dy
\left[\int d^4\theta K(\Phi^+,\Phi^-)
+\int d^2\theta^+W^+(\Phi^+)-\int d^2\theta^-W^-(\Phi^-)\right]
\nonumber\\
&=&S_K+S_W,
\label{N2S}
\eeqa
where $K(\Phi^+,\Phi^-)$ is a K\"ahler potential 
and $W^{\pm}(\Phi^{\pm})$ are superpotentials.
Consider first the case where only the K\"ahler potential term exists.
The supersymmetric variation of $S_K$ is
\beqa
\delta S_K&=&\int_{-\infty}^{\infty}dx\int_{-\infty}^0 dy\int d^4\theta
(\zeta^+Q_++\bar{\zeta}^+\bar{Q}_++\zeta^-Q_-+\bar{\zeta}^-\bar{Q}_-)
K(\Phi^+,\Phi^-)\qquad\qquad\ \nonumber\\
&=&\frac{i}{4}\int_{-\infty}^{\infty}dx
(\zeta^+K|_{\theta^+\bar{\theta}^+\bar{\theta}^-}
+\bar{\zeta}^+K|_{\theta^+\bar{\theta}^+\theta^-}
+\zeta^-K|_{\bar{\theta}^+\theta^-\bar{\theta}^-}
+\bar{\zeta}^-K|_{\theta^+\theta^-\bar{\theta}^-})|_{y=0}.
\label{dSK}
\eeqa
We can cancel (\ref{dSK}) by adding two types of boundary actions
\beq
S_{BK}=\frac{i}{4}\int_{-\infty}^{\infty}dx
\left(e^{i\beta}K|_{\bar{\theta}^+\theta^+}
+e^{-i\beta}K|_{\bar{\theta}^-\theta^-}\right),
\label{SBKA}
\eeq
or
\beq
S_{BK}=\frac{i}{4}\int_{-\infty}^{\infty}dx
\left(e^{i\beta}K|_{\theta^+\bar{\theta}^-}
+e^{-i\beta}K|_{\theta^-\bar{\theta}^+}\right),
\label{SBKB}
\eeq
where $e^{i\beta}$ is an arbirtary phase.
In the first case, the supersymmetry variation of $S_K+S_{BK}$ 
vanishes when $\bar{\zeta}^{\pm}=e^{\pm i\beta}\zeta^{\mp}$.
The conserved supercharges are
$Q_++e^{-i\beta}\bar{Q}_-$ and $Q_-+e^{i\beta}\bar{Q}_+$.
This leads to a condition  on the supercurrents:
$G^{\pm}+e^{\mp i\beta}\bar{G}^{\mp}=0$ at $y=0$.
This case is called as A-type boundary conditions \cite{OOY}.
The second case is $\bar{\zeta}^{\pm}=e^{\mp i\beta}\zeta^{\pm}$
where conserved supercharges are $Q_++e^{-i\beta}\bar{Q}_+$ 
and $Q_-+e^{i\beta}\bar{Q}_-$.
Associated boundary conditions on the supercurrents will be called as
B-type boundary condition:
$G^{\pm}+e^{\mp i\beta}\bar{G}^{\pm}=0$ at $y=0$.
In this paper, we will consider $e^{i\beta}=-1$
for simplicity.

With nonvanishing superpotential $W^{\pm}$,
the supersymmetric variation becomes
\beq
\delta S_W=\frac{1}{2}\int_{-\infty}^{\infty}dx
\left[(\bar{\zeta}^-\psi^--\zeta^-\bar{\psi}^-)
\frac{\partial W^+}{\partial\phi^+}
+(\zeta^+\bar{\psi}^+-\bar{\zeta}^+\psi^+)
\frac{\partial W^-}{\partial\phi^-}\right].
\label{dSW}
\eeq
We classify the boundary conditions into two classes following 
\cite{W,GJS}.

\noindent
{\bf A-type boundary condition}

We set $\bar{\zeta}^{\pm}=-\zeta^{\mp}$ in (\ref{dSW})
and assume that the fermions satisfy the condition
\beq
\psi^{\pm}-\bar{\psi}^{\mp}|_{y=0}=0.
\label{ABCpsi}
\eeq
The boundary conditions for the bosons are given by
\beq
\partial_x(\phi^+-\phi^-)=0,
~~~\partial_y(\phi^++\phi^-)=0.
\label{ABCphi}
\eeq
If the superpotentials $W^{\pm}$ satisfy
\beq
\frac{\partial W^+}{\partial\phi^+}-\frac{\partial W^-}
{\partial\phi^-}\Bigg|_{y=0}=0,
\label{ABCW}
\eeq
$\delta S_W=0$ can be achieved.

\noindent
{\bf B-type boundary condition}

If $\bar{\zeta}^{\pm}=-\zeta^{\pm}$, Eq.(\ref{dSW}) becomes
\beq
\delta S_W=\frac{1}{2}\int_{-\infty}^{\infty}dx
\left[-\zeta^-(\psi^-+\bar{\psi}^-)\frac{\partial W^+}{\partial\phi^+}
+\zeta^+(\psi^++\bar{\psi}^+)\frac{\partial W^-}{\partial\phi^-}\right].
\label{dSWB}
\eeq
This vanishes for two types of boundary conditions.

\noindent
(i) Dirichlet boundary conditions
\beq
\psi^{\pm}+\bar{\psi}^{\pm}|_{y=0}=0,
~~~\partial_x\phi^{\pm}|_{y=0}=0.
\label{BBCD}
\eeq

\noindent
(ii) Neumann boundary conditions
\beq
\psi^{\pm}-\bar{\psi}^{\pm}|_{y=0}=0,
~~~\partial_y\phi^{\pm}|_{y=0}=0.
\label{BBCN}
\eeq
While no additional condition is needed for (i),
the additional conditions
\beq
\frac{\partial W^{\pm}}{\partial\phi^{\pm}}\Bigg|_{y=0}=0
\label{BBCNW}
\eeq
are necessary for the case (ii).
To avoid this unphysical situation,
one must add additional boundary term
\beq
S_{BW}=\frac{i}{2}\eta\int_{-\infty}^{\infty}dx
(W^+-W^-)|_{\theta^{\pm}=\bar{\theta}^{\pm}=0}.
\label{SBW}
\eeq
The variation of this term cancels $\delta S_W$ in (\ref{dSWB})
if $\zeta^-+\eta\zeta^+=0$ is satsified.
This leads to the boundary conditions for $\phi^{\pm}$
\beq
\partial_y\phi^{\pm}\mp 2\pi i\eta\frac{\partial W^{\mp}}
{\partial\phi^{\mp}}\Bigg|_{y=0}=0.
\label{BBCNphi}
\eeq
Therefore, only $N=1$ supersymmetry is preserved.

\subsection{Boundary Action of $N=2$ Super-Liouville Theory}

Here we construct the boundary action with B-type boundary condition
which preserves the $N=2$ superconformal invariance.
We start with 
\beqa
S_B&=&\int_{-\infty}^{\infty}dx\Bigg[-\frac{i}{4\pi}
(\bar{\psi}^+\psi^-+\bar{\psi}^-\psi^+)
+\frac{1}{2}a^-\partial_x a^+
\nonumber\\
&&-\frac{1}{2}\left(f^+(\phi^+)a^++\tilde{f}^+(\phi^+)a^-\right)
(\psi^-+\bar{\psi}^-)
\nonumber\\
&&-\frac{1}{2}\left(f^-(\phi^-)a^-+\tilde{f}^-(\phi^-)a^+\right)
(\psi^++\bar{\psi}^+) +B(\phi^+,\phi^-)\Bigg],
\label{N2SLSB}
\eeqa
where $a^{\pm}$ are complex fermionic boundary degrees of freedom
and anti-commute with $\psi^{\pm}$ and $\bar{\psi}^{\pm}$.
The boundary action of the form (\ref{N2SLSB}) was first proposed
in the context of $N=2$ supersymmetric sine-Gordon model \cite{NN2}.
$f^{\pm}(\phi^{\pm})$, $\tilde{f}^{\pm}(\phi^{\pm})$ 
and $B(\phi^+,\phi^-)$ are functions of $\phi^{\pm}$ 
to be determined by the boundary conditions.

The stress tensor $T$, the supercurrent $G^{\pm}$ and 
the $U(1)$ current $J$ are given by
\beqa
&&T=-\partial\phi^-\partial\phi^+
-\frac{1}{2}(\psi^-\partial\psi^++\psi^+\partial\psi^-)
+\frac{1}{2}{\widehat Q}(\partial^2\phi^++\partial^2\phi^-),
\label{N2T}\\
&&G^{\pm}=\sqrt{2}i(\psi^{\pm}\partial\phi^{\pm}-{\widehat Q}
\partial\psi^{\pm}),
\label{N2G}\\
&&J=-\psi^-\psi^++{\widehat Q}(\partial\phi^+-\partial\phi^-),
\label{N2J}
\eeqa
where ${\widehat Q}$ is the background charge.

One can show that the conservation laws
$\bar{\partial}T=\partial\bar{T}
=\bar{\partial}G^{\pm}=\partial\bar{G}^{\pm}
=\bar{\partial}J=\partial\bar{J}=0$
are satisfied at the classical level when ${\widehat Q}=1/b$ 
in the same way as the $N=1$ case.
One major difference for the $N=2$ SLFT is
that ${\widehat Q}$ has no quantum correction.
Above conservation laws hold at quantum level with ${\widehat Q}=1/b$ due to
$:e^{b\phi^+}::e^{b\phi^-}:=:e^{b\phi^-}::e^{b\phi^+}:$.
This means that the classical level computation is sufficient 
for our consideration.
Also, without the correction, the dual symmetry $b\to 1/b$ disappears.
The lack of the dual symmetry makes it much harder to solve even bulk
$N=2$ SLFT \cite{AKRS}.

To preserve $N=2$ superconformal symmetry,
we impose the following boundary conditions on the conserved currents
\beq
T=\bar{T},
~~~G^{\pm}=\bar{G}^{\pm},
~~~J=\bar{J}
~~~{\rm at}
~~~y=0,
\label{N2bc}
\eeq

Substituting the bulk equations of motion 
\beq
\partial\bar{\partial}\phi^{\pm}
=\pi\mu b^3(i\psi^{\pm}\bar{\psi}^{\pm}+\pi\mu 
e^{b\phi^{\pm}})e^{b\phi^{\mp}}, \quad
\bar{\partial}\psi^{\pm}=-\pi i\mu b^2\bar{\psi}^{\mp}e^{b\phi^{\pm}},
\quad
\partial\bar{\psi}^{\pm}=\pi i\mu b^2\psi^{\mp} e^{b\phi^{\pm}}
\label{N2eom}
\eeq
and the boundary equations of motion 
\beqa
\partial_y\phi^{\pm}
&=&2\pi\left(\frac{\partial f^{\mp}}{\partial\phi^{\mp}}a^{\mp}
+\frac{\partial\tilde{f}^{\mp}}{\partial\phi^{\mp}}a^{\pm}\right)
(\psi^{\pm}+\bar{\psi}^{\pm})
-4\pi\frac{\partial B}{\partial\phi^{\mp}},
\nonumber\\
\psi^{\pm}-\bar{\psi}^{\pm}&=&-2\pi i(f^{\pm}a^{\pm}+\tilde{f}^{\pm}a^{\mp}),
\label{N2boeom}\\
\partial_x a^{\pm}&=&f^{\mp}(\psi^{\pm}+\bar{\psi}^{\pm})
+\tilde{f}^{\pm}(\psi^{\mp}+\bar{\psi}^{\mp})
\nonumber
\eeqa
into $G^{\pm}-\bar{G}^{\pm}$
and eliminating $\psi^{\pm}$ and $\bar{\psi}^{\pm}$, 
we obtain
\beqa
G^{\pm}-\bar{G}^{\pm}
&=&\pi\left(f^{\pm}-\frac{2}{b}\frac{\partial f^{\pm}}
{\partial\phi^{\pm}}\right) \partial_x\phi^{\pm}a^{\pm}
+\pi\left(\tilde{f}^{\pm}-\frac{2}{b}\frac{\partial\tilde{f}^{\pm}}
{\partial\phi^{\pm}}\right) \partial_x\phi^{\pm}a^{\mp}
\nonumber\\
&+&\pi\left(-\frac{2f^{\pm}}{f^{\pm}f^{\mp}-\tilde{f}^{\pm}\tilde{f}^{\mp}}
\frac{\partial B}{\partial\phi^{\mp}}
-\frac{2}{b}f^{\pm}
-\frac{\mu b\tilde{f}^{\mp}}{f^{\pm}f^{\mp}-\tilde{f}^{\pm}\tilde{f}^{\mp}}
e^{b\phi^{\pm}}\right)\partial_x a^{\pm}
\nonumber\\
&+&\pi\left(\frac{2\tilde{f}^{\pm}}{f^{\pm}f^{\mp}-\tilde{f}^{\pm}
\tilde{f}^{\mp}} \frac{\partial B}{\partial\phi^{\mp}}
-\frac{2}{b}\tilde{f}^{\pm}
+\frac{\mu bf^{\mp}}{f^{\pm}f^{\mp}-\tilde{f}^{\pm}\tilde{f}^{\mp}}
e^{b\phi^{\pm}}\right)\partial_x a^{\mp}.
\label{N2GbarG2}
\eeqa
The condition $G^{\pm}-\bar{G}^{\pm}=0$ determines
$f^{\pm}$, $\tilde{f}^{\pm}$ and $B$ as follows:
\beqa
&&f^{\pm}=C^{\pm}e^{b\phi^{\pm}/2},
~~~\tilde{f}^{\pm}=\tilde{C}^{\pm}e^{b\phi^{\pm}/2},
\label{N2f}\\
&&B=-\frac{2}{b^2}(C^+C^-+\tilde{C}^+\tilde{C}^-)e^{b(\phi^++\phi^-)/2},
\label{N2B}
\eeqa
where $C^{\pm}$ and $\tilde{C}^{\pm}$ are complex constants which obey
$C^{\pm}\tilde{C}^{\pm}=\mu b^2/4$.

We next consider the stress tensor.
Eliminating $a^{\pm}$ from (\ref{N2boeom})
and using (\ref{N2f}) and (\ref{N2B}), we obtain
\beqa
\partial_y\phi^{\pm}&=&i\frac{b}{2}(\psi^{\mp}-\bar{\psi}^{\mp})
(\psi^{\pm}+\bar{\psi}^{\pm})
+\frac{4\pi}{b}(C^+C^-+\tilde{C}^+\tilde{C}^-)e^{b(\phi^++\phi^-)/2},
\label{N2boeom4}\\
\partial_x\psi^{\pm}-\partial_x\bar{\psi}^{\pm}
&=&\frac{b}{2}\partial_x\phi^{\pm}(\psi^{\pm}-\bar{\psi}^{\pm})
-2\pi i(C^+C^-+\tilde{C}^+\tilde{C}^-)e^{b(\phi^++\phi^-)/2}
(\psi^{\pm}+\bar{\psi}^{\pm})
\nonumber\\
&-&4\pi iC^{\pm}\tilde{C}^{\pm}e^{b\phi^{\pm}}(\psi^{\mp}+\bar{\psi}^{\mp}).
\label{N2boeom5}
\eeqa
Substituting above equations into $T-\bar{T}$ and $J-\bar{J}$, one can
show that our solution satisfies both $T=\bar{T}$ and $J=\bar{J}$.

We have obtained the boundary action (\ref{N2SLSB})
with (\ref{N2f}) and (\ref{N2B}).
Moreover, we impose the invariance of ${\cal L}_B$ under 
the complex conjugation.
This invariance implies $C^+=(C^-)^*$
and $C^{\pm}$ can be written as $C^{\pm}=\mu_Be^{\pm i\alpha}$,
where $\alpha$ is a real parameter.
This phase factor can be gauged away
by redefining the fermionic zero-modes $a^{\pm}\to e^{\mp i\alpha}a^{\pm}$.
Therefore, the final form of the boundary action is
\beqa
S_B&=&\int_{-\infty}^{\infty}dx\Bigg[-\frac{i}{4\pi}
(\bar{\psi}^+\psi^-+\bar{\psi}^-\psi^+) +\frac{1}{2}a^-\partial_x a^+
\nonumber\\
&-&\frac{1}{2}e^{b\phi^+/2}\left(\mu_B a^+
+\frac{\mu b^2}{4\mu_B}a^-\right)(\psi^-+\bar{\psi}^-) \\
&-&\frac{1}{2}e^{b\phi^-/2}\left(\mu_B a^-
+\frac{\mu b^2}{4\mu_B} a^+\right)(\psi^++\bar{\psi}^+)
-\frac{2}{b^2}\left(\mu_B^2+\frac{\mu^2b^4}{16\mu_B^2}\right)
e^{b(\phi^++\phi^-)/2}\Bigg].\nonumber
\eeqa
This is our main result in this paper. This action preserves 
two conserved supercharges $Q_+-\bar{Q}_+$ and $Q_--\bar{Q}_-$.

We can rewrite this boundary action in terms of boundary superfields.
Defining super-translation operators $D_{t\pm}=D_{\pm}-\bar{D}_{\pm}$ 
which satisfy
\beq
\{D_{t+},D_{t-}\}=\partial_x, ~~~D_{t+}^2=D_{t-}^2=0,
\label{N2Dt}
\eeq
and their conjugate coordinates 
$\theta_t^{\pm}=(\theta^{\pm}-\bar{\theta}^{\pm})/2$,
we introduce fermionic boundary chiral superfields $\Gamma^{\pm}$ 
\beq
D_{t\mp}\Gamma^{\pm}=0.
\label{bchiral}
\eeq
The boundary superfields $\Gamma^{\pm}$ can be expanded as
\beq
\Gamma^{\pm}=a^{\pm}(x^{\pm})+i\theta_t^{\pm}h^{\pm}(x^{\pm}),
\label{bchiralex}
\eeq
where $x^{\pm}=x+\frac{1}{2}\theta_t^{\pm}\theta_t^{\mp}$.
In terms of these superfields, the boundary action can be written as
\beqa
&&S_B=\int_{-\infty}^{\infty}dx\Bigg\{-\frac{i}{4\pi}
\left(\Phi^+\Phi^-|_{\theta^+\bar{\theta}^-}
+\Phi^+\Phi^-|_{\theta^-\bar{\theta}^+}\right)
-\frac{1}{2}\mu(e^{b\Phi^+}+e^{b\Phi^-})|_{\theta^{\pm}=\bar{\theta}^{\pm}=0}
\nonumber\\
&+&\frac{1}{2}\int d\theta_t^+d\theta_t^-\Gamma^+\Gamma^-
\nonumber\\
&-&\Bigg[{i\over{b}}\int d\theta_t^+\left(\mu_B\Gamma^+e^{b\Phi^+/2}
+\frac{\mu b^2}{4\mu_B}\left(\Gamma^-e^{b\Phi^+/2}
-\Gamma^+e^{b\Phi^-/2}\right)\right)\Bigg|_{\theta_t^-=0}
+{\rm c.c.}\Bigg]\Bigg\}.
\label{N2SLSBsuper}
\eeqa
When the terms including the superfields $\Gamma^{\pm}$ do not exist,
(\ref{N2SLSBsuper}) reduces to the boundary action
which preserves only $N=1$ supersymmetry under Neumann boundary conditions.
In this case the $N=2$ supersymmetry transformation of the action 
$S+S_B$ has a non-vanishing surface term
which is canceled by those of the terms including $\Gamma^{\pm}$.

\section{Discussions}

Our result contains one boundary parameter $\mu_B$
which generates a continuous family of conformal boundary conditions.
One remarkable result is that the boundary action has a dual symmetry
\beq
\mu_B\to \frac{\mu b^2}{4\mu_B}.
\label{dual}
\eeq
This means that two conformal boundary conditions of the $N=2$ SLFT
can be identified.
To understand further implications about this, we need to derive 
some exact correlation functions such as the boundary one-point functions.
Our boundary action is a first step toward this.
It is possible to derive a functional relation for the one-point
functions using the boundary action as a screening boundary operator.
Main difficulty arises, as in the bulk case \cite{AKRS}, from the
lack of the coupling constant duality.
In a recent paper \cite{MMV}, the one-point functions for the 
$N=2$ SLFT is conjectured from the modular transformations of
the characters for a special value of coupling constant.
It would be interesting to see if these one-point functions are
consistent with the functional relations based on our boundary action
and to derive them for arbitary value of the coupling constant.

\section*{Acknowledgments}

We thank C. Kim, R. Nepomechie, J. Park, C. Rim and M. Stanishkov
for helpful discussions.
This work was supported in part by Korea Research Foundation 
2002-070-C00025.

\appendix

\section{Conventions}
In this appendix we present our conventions for $N=1,2$  supersymmetries.

\subsection{$N=1$ Supersymmetry}

We use $(1,1)$ superspace with bosonic coordinates $z$, $\bar{z}$
and fermionic coordinates $\theta$, $\bar{\theta}$.
Here we define $z=x+iy$, $\bar{z}=x-iy$
and $\partial=(\partial_x-i\partial_y)/2$, 
$\bar{\partial}=(\partial_x+i\partial_y)/2$.
The integration measure is $\int d^2zd^2\theta=\int dxdyd\theta d\bar{\theta}$.
The covariant derivatives are given by
\beq
D=\frac{\partial}{\partial\theta}+\theta\partial,
~~~\bar{D}=\frac{\partial}{\partial{\bar{\theta}}}+\bar{\theta}\bar{\partial}.
\label{N1D}
\eeq
which satisfy
\beq
\{D,D\}=2\partial,
~~~\{\bar{D},\bar{D}\}=2\bar{\partial},
~~~\{D,\bar{D}\}=0.
\label{N1Dalgebra}
\eeq
The supercharges 
\beq
Q=\frac{\partial}{\partial\theta}-\theta\partial,
~~~\bar{Q}=\frac{\partial}{\partial\bar{\theta}}-\bar{\theta}\bar{\partial}.
\label{N1Q}
\eeq
satisfy
\beq
\{Q,Q\}=-2\partial,
~~~\{\bar{Q},\bar{Q}\}=-2\bar{\partial},
~~~\{Q,\bar{Q}\}=0,
\label{N1Qalgebra}
\eeq
and anti-commute with $D$, $\bar{D}$.

\subsection{$N=2$ Supersymmetry}

We use $(2,2)$ superspace with bosonic coordinates $z$, $\bar{z}$
and fermionic ones $\theta^+$, $\bar{\theta}^+$, $\theta^-$, $\bar{\theta}^-$.
Complex conjugation of fermionic coordinates is defined by 
$(\theta^{\pm})^*=\bar{\theta}^{\mp}$.
The covariant derivatives 
\beq
D_{\pm}=\frac{\partial}{\partial\theta^{\pm}}+\frac{1}{2}\theta^{\mp}\partial,
~~~\bar{D}_{\pm}=\frac{\partial}{\partial\bar{\theta}^{\pm}}
+\frac{1}{2}\bar{\theta}^{\mp}\bar{\partial}.
\label{N2D}
\eeq
satisfy
\beqa
&&\{D_+,D_-\}=\partial,
~~~\{\bar{D}_+,\bar{D}_-\}=\bar{\partial},
\nonumber\\
&&\mbox{all other (anti-)commutators}=0.
\label{N2Dalgebra}
\eeqa
The supercharges are given by
\beq
Q_{\pm}=\frac{\partial}{\partial\theta^{\pm}}-\frac{1}{2}\theta^{\mp}\partial,
~~~\bar{Q}_{\pm}=\frac{\partial}{\partial\bar{\theta}^{\pm}}
-\frac{1}{2}\bar{\theta}^{\mp}\bar{\partial},
\label{N2Q}
\eeq
which obey
\beqa
&&\{Q_+,Q_-\}=-\partial,
~~~\{\bar{Q}_+,\bar{Q}_-\}=-\bar{\partial},
\nonumber\\
&&\mbox{all other (anti-)commutators}=0,
\label{N2Qalgebra}
\eeqa
and anti-commute with $D_{\pm}$, $\bar{D}_{\pm}$.
Chiral superfields $\Phi^{\pm}_i$ satisfy 
$D_{\mp}\Phi^{\pm}_i=\bar{D}_{\mp}\Phi^{\pm}_i=0$.
An $N=2$ supersymmetric action is constructed from D-terms and 
F-terms\footnote{One can also consider
twisted F-terms which we do not mention here.}
and is written as
\beq
S=\int d^2zd^4\theta K(\Phi^+_i,\Phi^-_i)
+\left(\int d^2zd^2\theta^+W^+(\Phi^+_i)+{\rm c.c.}\right),
\label{N2SKW}
\eeq
where $K(\Phi^+_i,\Phi^-_i)$ is an arbitrary differentiable function of 
superfields and $W^+(\Phi^+_i)$ is a holomorphic function of 
chiral superfields $\Phi^+_i$.
The integration measures in (\ref{N2SKW}) are defined by
\beqa
&&\int d^2zd^4\theta K(\Phi^+_i,\Phi^-_i)
=\int dxdyd\theta^+d\bar{\theta}^+d\theta^-d\bar{\theta}^-
K(\Phi^+_i,\Phi^-_i), \label{N2K}\\
&&\int d^2zd^2\theta^+W^+(\Phi^+_i)
=\int dxdyd\theta^+d\bar{\theta}^+W^+(\Phi^+_i)
\Big|_{\theta^-=\bar{\theta}^-=0}.
\label{N2W}
\eeqa

\section{Quantum Corrections of the $N=1$ boundary supersymmetry}

In this appendix we show how the classical boundary action (\ref{N1SLSB})
with Eqs.(\ref{N1B}) is modified at the quantum level.
At the quantum level, we replace $e^{\alpha\phi}$ 
with the normal-ordered exponential $:e^{\alpha\phi}:$.
Therefore, we can not elliminate $\psi,\bar{\psi}$ from $G,\bar{G}$
because $f^{-1}$ may not be well-defined.
Therefore, we need to keep the fermionic fields.
Then, Eq.(\ref{N1GbarG}) becomes
\beqa
G-\bar{G}&=&2\pi fa\partial_x\phi+2\pi(\psi+\bar{\psi})\frac{\partial f}{\partial\phi}a(\psi+\bar{\psi})
-2\pi(\psi+\bar{\psi})\frac{\partial B}{\partial\phi}
\nonumber\\
&-&4\pi{\widehat Q}\partial_x fa-4\pi{\widehat Q}f^2(\psi+\bar{\psi})
+{\widehat Q}\pi\mu b^2(\psi+\bar{\psi})\Lambda^{-2b^2}(:e^{b\phi/2}:)^2,
\label{N1GbarGq1}
\eeqa
where $\Lambda$ is a cutoff scale and $f=\mu_B:e^{b\phi/2}:$.

The first and second terms in the right-hand side of Eq.(\ref{N1GbarGq1})
can be dealt with the point-splitting technique developed in \cite{LM}.
The first term can be calculated as
\beqa
fa\partial_x\phi&=&\frac{\mu_B}{2}\lim_{x_1\to x_2}
[:e^{b\phi(x_1)/2}:a(x_1)\partial_x\phi(x_2) +(x_1\leftrightarrow x_2)]
\nonumber\\
&=&\mu_B:e^{b\phi/2}\partial_x\phi:a+b\lim_{x_1\to x_2}
\frac{1}{x_1-x_2}[f(x_1)a(x_1)-f(x_2)a(x_2)] \nonumber\\
&=&\frac{2}{b}\partial_x fa+b\partial_x fa+bf^2(\psi+\bar{\psi}),
\label{N1GbarGqpst1}
\eeqa
where we used (\ref{N1boeom}).
Similarly the second term becomes
\beq
(\psi+\bar{\psi})fa(\psi+\bar{\psi})=2\partial_x fa+2f^2(\psi+\bar{\psi}).
\label{N1GbarGqpst2}
\eeq
This leads to
\beq
G-\bar{G}=\pi(\psi+\bar{\psi})\left[\left(-\frac{4}{b}\mu_B^2+{\widehat Q}
\mu b^2\Lambda^{-2b^2}\right)(:e^{b\phi/2}:)^2 -2\frac{\partial B}{\partial\phi}\right],
\label{N1GbarGq2}
\eeq
where we used ${\widehat Q}=1/b+b$.
The condition $G-\bar{G}=0$ gives
\beq
B=\left(-\frac{2}{b^2}\mu_B^2+\frac{1}{2}{\widehat Q}b\mu\Lambda^{-2b^2}\right)(:e^{b\phi/2}:)^2.
\label{N1Bq}
\eeq
Compared with the classical result (\ref{N1B}),
$B$ gets the quantum correction.

The condition $T-\bar{T}=0$ can be also satisfied in this way.
Substituting Eqs.(\ref{N1eom}) and (\ref{N1boeom}) into $T-\bar{T}$,
we obtain
\beq
T-\bar{T}=\pi i[2\partial_x fa(\psi+\bar{\psi})
-\partial_x(fa(\psi+\bar{\psi}))]
+\frac{1}{2}(\bar{\psi}\partial_x\bar{\psi}-\psi\partial_x\psi).
\label{N1TbarTq2}
\eeq
Using Eq.(\ref{N1boeom}) for $\psi$ and $\bar{\psi}$, one can show that
Eq.(\ref{N1TbarTq2}) vanishes.
Therefore, the boundary action (\ref{N1SLSB}) along with $f$ and $B$ given 
above preserves the boundary conformal symmetry upto the quantum level.

\end{document}